%% file: main.tex
\documentclass[sigconf,natbib=true,screen=true]{acmart}

\acmSubmissionID{4465}

\input{packages}

\input{definitions}

\input{authors}

\input{meta}

\begin{document}

\title[Going Beyond Popularity and Positivity Bias: Correcting for Multifactorial Bias in Recommender Systems]{
Going Beyond Popularity and Positivity Bias:\\
Correcting for Multifactorial Bias in Recommender Systems}

\begin{abstract}
Two typical forms of bias in user interaction data with \acp{RS} are popularity bias and positivity bias,
which manifest themselves as the over-representation of interactions with popular items or items that users prefer, respectively.
Debiasing methods aim to mitigate the effect of selection bias on the evaluation and optimization of \acp{RS}.
However, existing debiasing methods only consider single-factor forms of bias, \eg only the item (popularity) or only the rating value (positivity).
This is in stark contrast with the real world where user selections are generally affected by multiple factors at once.
In this work, we consider multifactorial selection bias in \acp{RS}.
Our focus is on selection bias affected by both item and rating value factors, which is a generalization and combination of popularity and positivity bias.
While the concept of multifactorial bias is intuitive, it brings a severe practical challenge as it requires substantially more data for accurate bias estimation.
As a solution, we propose smoothing and alternating gradient descent techniques to reduce variance and improve the robustness of its optimization.
Our experimental results reveal that, with our proposed techniques, multifactorial bias corrections are more effective and robust than single-factor counterparts on real-world and synthetic datasets.
\end{abstract}

\maketitle

\acresetall

\input{sections/sec-intro}

\input{sections/sec-concept}

\input{sections/sec-preliminaries-merged}

\input{sections/sec-model}

\input{sections/sec-experiment}

\input{sections/sec-biases-effect}

\input{sections/sec-rel}

\input{sections/sec-con}
\input{sections/acknowledge}

\clearpage
\bibliographystyle{ACM-Reference-Format}
\balance
\bibliography{references}

\end{document}

%% file: packages.tex
\usepackage{xcolor}
\usepackage{graphicx}
\usepackage[skip=1pt]{caption}
\usepackage{subcaption}
\usepackage{bm}
\usepackage{bbm}
\usepackage{dsfont}
\usepackage{xspace}
\usepackage{url}
\usepackage{multirow}
\usepackage{booktabs}
\usepackage{enumerate}
\usepackage[inline]{enumitem}
\usepackage{acronym}
\usepackage{balance}
\usepackage{mleftright}

\usepackage[linesnumbered,ruled]{algorithm2e}
\usepackage{tikz}
\usetikzlibrary{arrows.meta}

\usepackage{makecell}

%% file: definitions.tex
\newcommand{\ie}{\emph{i.e.,}\xspace}

\newcommand{\eg}{\emph{e.g.,}\xspace}

\newcommand{\ignore}[1]{}

\definecolor{HarrieGreen}{RGB}{34,139,34}

\acrodef{RL}{reinforcement learning}
\acrodef{RL4Rec}{reinforcement learning for recommendation}
\acrodef{ID}{in-distribution}
\acrodef{OOD}{out-of-distribution}
\acrodef{Seq-Rec}{sequential recommendation}
\acrodef{RS}{recommender system}
\acrodef{EIB}{error-imputation-{\allowbreak}based model}
\acrodef{IPS}{inverse propensity scoring}
\acrodef{PWE}{propensity-weighted estimator}
\acrodef{DR}{doubly robust}
\acrodef{SNIPS}{self-normalized inverse propensity scoring}
\acrodef{MNAR}{missing not at random}
\acrodef{Pop}{popularity}
\acrodef{Avg}{average}
\acrodef{MF}{matrix factorization}
\acrodef{BPR}{Bayesian personalized ranking}
\acrodef{TMF}{time-aware matrix factorization}
\acrodef{TTF}{time-aware tensor factorization}
\acrodef{TMTF}{time-aware matrix \& tensor factorization}
\acrodef{MC}{Markov chains}
\acrodef{DQN}{Deep Q-Network}
\acrodef{DQN4Rec}{Deep Q-Network based recommendation}
\acrodef{MDP}{Markov decision process}
\acrodef{GT}{Ground Truth}
\acrodef{sim-GT}{simulated Ground Truth}
\acrodef{MCAR}{missing completely at random}
\acrodef{CF}{collaborative filtering}
\acrodef{RNN}{recurrent neural network}
\acrodef{GRU}{gated recurrent unit}
\acrodef{CNN}{convolutional neural network}
\acrodef{MLP}{multi-layer perceptron}
\acrodef{LSTM}{long short-term memory}
\acrodef{BOI}{bag of items}
\acrodef{PLD}{pairwise local dependency between items}

\acrodef{NLL}{Negative Log-Likelihood}
\acrodef{PPL}{Perplexity}

\acrodef{P}{Precision}
\acrodef{R}{Recall}
\acrodef{MAP}{Mean Average Precision}
\acrodef{MRR}{Mean Reciprocal Rank}
\acrodef{NDCG}{normalized discounted cumulative gain}

\acrodef{MSE}{mean squared error}
\acrodef{MAE}{mean absolute error}
\acrodef{RMSE}{root mean square error} 
\acrodef{ALS}{alternating least squares}

\makeatletter
\renewcommand\paragraph{\@startsection{paragraph}{4}{\z@}%
                       {-2\p@ \@plus -1\p@ \@minus -1\p@}%
                       {-0.5em \@plus -0.22em \@minus -0.1em}%
                       {\normalfont\normalsize\itshape}}
\makeatother

\SetCommentSty{mycommfont}

\newcommand{\header}[1]{\vspace*{1mm}\noindent\textbf{#1}.}

%% file: authors.tex
\author{Jin Huang}
\orcid{0000-0001-9273-9037}
\affiliation{%
\institution{University of Amsterdam}
  \city{Amsterdam}
  \country{The Netherlands}
}
\email{j.huang2@uva.nl}

\author{Harrie Oosterhuis}
\orcid{0000-0002-0458-9233}
\affiliation{%
\institution{Radboud University}
  \city{Nijmegen}
  \country{The Netherlands}
}
\email{harrie.oosterhuis@ru.nl}

\author{Masoud Mansoury}
\orcid{0000-0002-9938-0212}
\affiliation{%
\institution{Delft University of Technology}
  \city{Delft}
  \country{The Netherlands}
}
\email{m.mansoury@tudelft.nl}
\authornote{Work done while the author was with Elsevier Discovery Lab and University of Amsterdam.}

\author{Herke van Hoof}
\orcid{0000-0002-1583-3692}
\affiliation{%
\institution{University of Amsterdam}
  \city{Amsterdam}
  \country{The Netherlands}
}
\email{h.c.vanhoof@uva.nl}

\author{Maarten de Rijke}
\orcid{0000-0002-1086-0202}
\affiliation{%
\institution{University of Amsterdam}
  \city{Amsterdam}
  \country{The Netherlands}
}
\email{m.derijke@uva.nl}

\renewcommand{\shortauthors}{J. Huang et al.}

\def\authors{Jin Huang, Harrie Oosterhuis, Masoud Mansoury, Herke van Hoof, and\linebreak Maarten de Rijke}

%% file: meta.tex
\copyrightyear{2024}
\acmYear{2024}
\setcopyright{rightsretained}
\acmConference[SIGIR '24]{Proceedings of the 47th International ACM SIGIR Conference on Research and Development in Information Retrieval}{July 14--18, 2024}{Washington, DC, USA}
\acmBooktitle{Proceedings of the 47th International ACM SIGIR Conference on Research and Development in Information Retrieval (SIGIR '24), July 14--18, 2024, Washington, DC, USA}
\acmDOI{10.1145/3626772.3657749}
\acmISBN{979-8-4007-0431-4/24/07}

\ccsdesc[500]{Information systems~Recommender systems}

\keywords{Recommender Systems; Unbiased Learning; Propensity Estimation}

%% file: sections/sec-intro.tex
\vspace*{-2mm}
\section{Introduction}
\label{section:introduction}
Rating prediction is a fundamental \ac{RS} task where the goal is to predict user ratings on items.
The task facilitates personalized recommendations to improve user satisfaction~\cite{bobadilla2013recommender,ricci2010introduction,ricci2015recommender}.
Rating prediction methods that are learned from user ratings can be biased as user interactions with \acp{RS} are subject to severe selection bias~\cite{schnabel2016recommendations, ovaisi2020correcting, marlin2009collaborative, pradel2012ranking,steck2011item}.
The effects of such bias can produce systematic errors in user preference prediction~\citep{huang2020keeping,schnabel2016recommendations,yao2021measuring} and result in problems of over-specialization~\cite{adamopoulos2014over}, filter bubbles~\cite{nguyen2014exploring, pariser2011filter}, and unfairness~\cite{chen2023bias}.
Two influential types of bias present in user rating behavior are popularity bias~\citep{pradel2012ranking, steck2011item, canamares2018should} and positivity bias~\cite{pradel2012ranking}, which arise as users are more likely to rate popular items or items that they prefer, respectively.

\header{Single-factor bias}
Widely-used methods for mitigating the effect of selection bias in user ratings make use of \ac{IPS}~\cite{imbens2015causal} and integrate it into the learning process~\cite{schnabel2016recommendations, huang2020keeping, joachims2017unbiased}.
Given the propensity of a rating, \ie the probability of the corresponding user rating the specific item, \ac{IPS} weights each rating inversely to their propensity, and, thereby, corrects for the over-representation resulting from selection bias.
The predominant model of popularity bias in previous work assumes that the propensity values only depend on the corresponding item.
For positivity bias, the propensity values are assumed to only depend on the corresponding rating value. 
These single-factor propensity models can provide unbiased estimations with \ac{IPS}, given that their assumptions about the factors that determine the selection bias in user data are correct.
However, real-world user decisions about rating items generally depend on more than one factor, a scenario 
that existing methods do not address in practice~\cite{pradel2012ranking,idrissiguess,ekstrand2018all}.

\header{Multifactorial bias}
We consider a \emph{multifactorial} bias that is determined by two factors, \ie item and rating value.
This can be seen as a generalization of popularity and positivity bias that combines the essential properties of both.
As we expect multifactorial bias to better capture actual user behavior, we also expect that the resulting propensities will lead to a better performance of \ac{IPS}-based debiasing methods.
To estimate multifactorial bias, existing propensity estimation methods~\cite{schnabel2016recommendations}, based on naive Bayes or logistic regression, can simply be used by accommodating multiple factors.
Surprisingly, there is a lack of studies comparing the performance of \ac{IPS}-based debiasing methods using single-factor bias estimation against those using such a multifactorial bias estimation.
This raises questions about the practical utility of multifactorial bias estimation and correction.
Moreover, our experimental results on real-world datasets indicate that existing multifactorial bias estimation methods lead to unstable performance when applied to \ac{IPS}-based debiasing methods.
This could potentially explain their limited adoption in practice.

The practical challenges associated with multifactorial bias arise as the consideration of multiple factors greatly increases problems of data sparsity~\cite{ricci2010introduction,da2018effects}.
For comparison, single-factor popularity bias estimation is based on the observation frequency of ratings per item, \ie how many users have rated an item.
Single-factor positivity bias estimation is based on the difference in frequency of rating values between naturally observed ratings and a (small) unbiased dataset, \ie how much more often or less often a rating value is observed in natural user interactions than when users rate randomly sampled items.
Both single-factor estimation techniques already have to deal with severe sparsity, as most items are not very popular and often only very little unbiased data is available~\cite{ricci2010introduction,da2018effects,burke2002hybrid}.
Multifactorial bias estimation exacerbates this sparsity problem as it has to consider the frequencies of combinations of items and rating values.
As a result, before a multifactorial bias approach can be effective, one has to first overcome this severe data-efficiency problem.

\header{Contributions and findings}
In this work, we develop a propensity estimation method for multifactorial bias that is determined by item and rating value factors. 
The results of our proposed multifactorial bias propensity estimation are integrated with an \ac{IPS}-based debiasing method to correct for multifactorial bias.
To deal with the severe sparsity problem multifactorial bias poses, we propose the adoption of propensity smoothing technique and an alternating gradient descent approach for more robust and stable \ac{IPS}-based optimization.

To evaluate our multifactorial method, we compare the \ac{IPS}-based debiasing method using our multifactorial bias estimation against those using single-factor bias estimation on a selection of real-world datasets: the \text{Yahoo!R3}~\cite{marlin2009collaborative}, Coat~\cite{schnabel2016recommendations}, and KuaiRec~\cite{gao2022kuairec} datasets.
Our experimental results show the effectiveness of our multifactorial method over state-of-the-art single-factor counterparts.
Furthermore, we perform an extensive simulation-based experimental analysis where the effect of each of the two factors is varied.
The results show that single-factor methods are only effective when their corresponding factor dominates selection bias, but perform poorly when the other factor is also important.
In contrast, our multifactorial approach has much more robust performance, as it is always effective, regardless of how much effect each factor has, and provides considerably better performance when both factors have a substantial effect.
This indicates that, once its sparsity problem is dealt with, our multifactorial approach provides the safest choice when it is unclear what factors determine selection bias.

%% file: sections/sec-concept.tex
\section{Conceptualization of selection bias}
\label{sec:bias-conceptualization}
In this section, we provide an overview of existing conceptualizations of selection bias, popularity bias, and positivity bias in the context of \acp{RS}.
Some concepts have seen varied definitions across publications, potentially leading to confusion in their usage. 

\header{Selection bias}
\citet{ovaisi2020correcting} conclude that selection bias occurs when a data sample is not representative of the underlying data distribution.
Primary studies delineate selection bias into two principal categories~\cite{saito2020asymmetric,schnabel2016recommendations,wang2019doubly,huang2022different}: self-selection bias, where users choose to interact with certain items more often, and algorithmic bias, where items showing to users are highly dependent on the algorithm in an \ac{RS}.
In this paper, we adopt a definition of selection bias in line with~\citet{ovaisi2020correcting}.

In some studies, the definition of selection bias slightly differs.
\citet{chen2023bias} constrain selection bias exclusively to self-selection bias while delineating algorithmic bias as exposure bias.
Exposure bias could be known as ``previous model bias'' when the previous recommendation policy controls what items to show~\cite{liu2020general}, or ``user-selection bias'' in the scenario where the \ac{RS} shows the items according to users' active search queries~\cite{wang2016learning}.

\header{Popularity bias}
Prior work mostly takes popularity to be a form of selection bias defined by~\citet{ovaisi2020correcting} or exposure bias defined by~\citet{chen2023bias}.
Popularity bias is often defined based on two primary reasons for occurrence: users are more likely to provide feedback on popular items~\cite{steck2011item,pradel2012ranking}, and popular items are recommended more frequently than their popularity would warrant~\cite{abdollahpouri2020multi,mansoury2020feedback,zhu2021popularity,celma2008hits}.
Due to popularity bias, the observed logged data reveals a concentration of user interactions on popular items, shown as a long-tail distribution in the frequency of interactions across items.
Therefore, there exists a widely shared consensus that popularity bias is closely associated with the long-tail phenomenon~\cite{steck2011item,abdollahpouri2020multi,chen2023bias}.

\header{Positivity bias}
Positivity bias is another form of selection bias, which is uniformly considered to refer to the scenario where users rate more often the items they like~\cite{pradel2012ranking,park2018positivity}.
In contrast, a rarely studied but relevant form of selection bias could occur when users rate more often the items they dislike.\footnote{We avoid the use of the term negativity bias, as it commonly denotes a scenario where wrong impressions may sometimes outweigh good ones~\cite{baumeister2001bad}.}
These forms of bias can contribute to a scenario where observed user ratings are characterized by a skewed rating distribution compared to the true rating distribution~\cite{pradel2012ranking,huang2020keeping}.

%% file: sections/sec-preliminaries-merged.tex
\section{Preliminaries}
\label{sec:bias-defintions}

Before we define and address multifactorial bias specifically, we introduce our problem setting, provide a formal definition of selection bias, and summarize the \ac{IPS}-based debiasing methods.

We follow the common \ac{RS} setting where users from a set $\mathcal{U}=\{u_1,\ldots, u_N\}$ give ratings on items from a set $\mathcal{I}=\{i_1, \ldots, i_M \}$~\cite{steck2013evaluation}. 
User preferences are explicitly shown by these ratings, $y_{u,i} \in \mathcal{R} = \{1,2, 3,4,5 \}$ per user $u \in \mathcal{U}$ and item $i \in \mathcal{I}$.
In practice, logged rating data $\mathcal{D}$ is often very sparse and subject to heavy selection bias as it is unrealistic for all users to provide ratings for all items.
To indicate which ratings are available for optimization, we use an observation indicator matrix $\bm{O} \in \{0,1\}^{|\mathcal{U}| |\mathcal{I}|}$, where $o_{u,i} \in \bm{O}$ indicates whether the rating for user $u$ on item $i$ is recorded in the logged data ($o_{u,i}=1$) or not ($o_{u,i}=0$).
One can expect $\bm{O}$ to be sparse and influenced by selection bias~\cite{schnabel2016recommendations,steck2011item,huang2020keeping}.
Next, we define several forms of selection bias and discuss their effects on rating prediction methods that learn from logged rating data: $\mathcal{D}=\{(u,i,y_{u,i}) \mid u\in \mathcal{U}, i\in \mathcal{I}, o_{u,i}=1\}$.

\subsection{Definition of selection bias}

As discussed in Section~\ref{sec:bias-conceptualization}, selection bias occurs if the process that decides whether a user rates an item is not a random selection.
We formally define selection bias by using
the propensities $p_{u,i}$, \ie the probabilities of a user rating an item: $p_{u,i} = P(o_{u,i}=1 \mid u,i, y_{u,i})$.

\begin{definition}[Selection bias]
    Logged rating data $\mathcal{D}$ is subject to \emph{selection bias} if not every rating propensity has the same value:
    \begin{equation}
        \begin{split}
            \textit{Selection-bias}(\mathcal{D}) \Longleftrightarrow 
            \exists u, u'\in \mathcal{U}\,, \exists i, i' \in \mathcal{I}, \; p_{u,i} \not= p_{u',i'}.
        \end{split}
    \end{equation}
\end{definition}

\begin{figure}[t]
    \centering
    \begin{subfigure}{0.3\linewidth}
        \centering
        \begin{tikzpicture}
            \node[circle, draw, minimum size=0.2cm] (circle1) at (0,0) {I};
            \node[circle, draw, minimum size=0.2cm] (circle2) at (1.5,0) {Y};
            \node[circle, draw, minimum size=0.2cm] (circle3) at (0.75,-1.2) {O};
            \draw[-{Latex[length=2mm]}, line width=0.5pt] (circle1) -- (circle2);
            \draw[-{Latex[length=2mm]}, line width=0.5pt] (circle2) -- (circle3);
        \end{tikzpicture}
        \caption{}
        \label{fig:sub:pos-bias} 
    \end{subfigure}
  \hfill
    \begin{subfigure}{0.3\linewidth}
        \centering
      \begin{tikzpicture}
        \node[circle, draw, minimum size=0.2cm] (circle1) at (0,0) {I};
        \node[circle, draw, minimum size=0.2cm] (circle2) at (1.5,0) {Y};
        \node[circle, draw, minimum size=0.2cm] (circle3) at (0.75,-1.2) {O};
        \draw[-{Latex[length=2mm]}, line width=0.5pt] (circle1) -- (circle2);
        \draw[-{Latex[length=2mm]}, line width=0.5pt] (circle1) -- (circle3);
      \end{tikzpicture}
    \caption{}
    \label{fig:sub:pop-bias}
    \end{subfigure}
  \hfill
    \begin{subfigure}{0.3\linewidth}
        \centering
      \begin{tikzpicture}
        \node[circle, draw, minimum size=0.2cm] (circle1) at (0,0) {I};
        \node[circle, draw, minimum size=0.2cm] (circle2) at (1.5,0) {Y};
        \node[circle, draw, minimum size=0.2cm] (circle3) at (0.75,-1.2) {O};
        \draw[-{Latex[length=2mm]}, line width=0.5pt] (circle1) -- (circle2);
        \draw[-{Latex[length=2mm]}, line width=0.5pt] (circle2) -- (circle3);
        \draw[-{Latex[length=2mm]}, line width=0.5pt] (circle1) -- (circle3);
      \end{tikzpicture}
      \caption{}
    \label{fig:sub:mul-bias}
    \end{subfigure}
    \caption{The dependency between observance (O), items (I), and rating values (Y) for different bias assumptions: 
    (a) Positivity bias: propensities only depend on rating values; 
    (b) Popularity bias: propensities only depend on items; 
    (c) Multifactorial bias: propensities depend on both factors. %
    }
    \label{fig:bias-definitions}
  \end{figure}
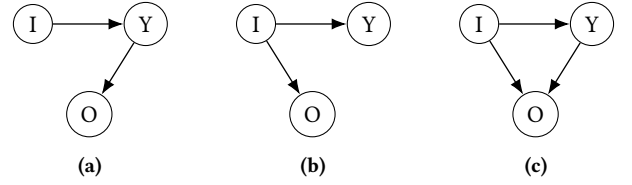

\noindent%
We further provide the following definitions of two influential forms of selection bias -- positivity bias and popularity bias -- to match our usage of the terms:

\begin{definition}[Positivity bias]
    \label{def:pos-bias}
    Logged rating data $\mathcal{D}$ is subject to \emph{positivity bias} if propensities  only depend on their rating values (Fig.~\ref{fig:sub:pos-bias}) and higher ratings correspond to higher propensities:
        \begin{equation}
        \begin{split}
             &\textit{Positivity-bias}(\mathcal{D}) \Longleftrightarrow 
             \big(\textit{Selection-bias}(\mathcal{D}) \, \land \\
            & \hspace{0.7cm} \forall u, u' \in \mathcal{U}, \forall i, i' \in \mathcal{I}, \;
             \big( y_{u,i} > y_{u', i'} \longleftrightarrow p_{u,i} > p_{u',i'} \big)\big).
        \end{split}
    \end{equation}
\end{definition}

\begin{definition}[Popularity bias]
    \label{def:pop-bias}
    Logged rating data $\mathcal{D}$ is subject to \emph{popularity bias} if the propensities of ratings only depend on which item they correspond to (Fig.~\ref{fig:sub:pop-bias}):
    \begin{equation}
        \begin{split}
            & \textit{Popularity-bias}(\mathcal{D}) \Longleftrightarrow \big(\textit{Selection-bias}(\mathcal{D}) \, \land \\
            & \hspace{1.65cm} \forall u, u'\in\mathcal{U}, \forall i, i'\in\mathcal{I},
           \; \big(i = i' \longrightarrow p_{u,i} = p_{u',i'} \big) \big).
        \end{split}
    \end{equation}
\end{definition}

\begin{figure*}[t]
    \centering
    \hspace{-16pt}
    \begin{subfigure}{0.25\linewidth}
    \raisebox{10pt}{
    \includegraphics[clip, trim=0mm 0mm 0mm 0mm, scale=0.75]{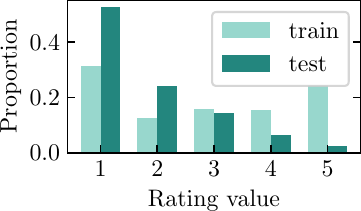}
    }
    \caption{Distribution of rating values.}
    \label{fig:yahoo:rating}
    \end{subfigure}
    \hspace{5pt}
    \begin{subfigure}{0.30\textwidth}
    \raisebox{10pt}{
    \includegraphics[clip, trim=0mm 0mm 0mm 0mm, scale=0.75]{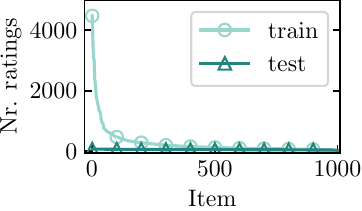}
    }
    \caption{Distributions of interactions over items.}
    \label{fig:yahoo:itempopularity}
    \end{subfigure}
    \begin{subfigure}{0.40\linewidth}
        \includegraphics[clip, trim=0mm 5.8mm 0mm 0mm, scale=0.58]{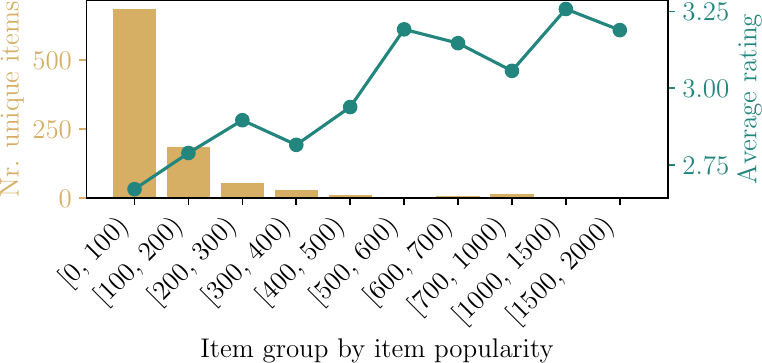}
    \vspace{-17pt}
    \caption{Item group by item popularity.}
    \label{fig:correlated_rating_item}
    \end{subfigure}
    \caption{Skewed distributions of (a) rating values or (b) item popularity in the logged training set (train) of the \text{Yahoo!R3} dataset, and (c) the number and average ratings of items in a group that contains items with the number of interactions falling within a certain interval are counted from
    logged user ratings on the self-selected songs in the Yahoo!R3 dataset.} 
    \vspace{-12pt}
    \label{fig:yahoo}
\end{figure*}

\noindent%
As discussed in Section~\ref{sec:bias-conceptualization}, the definition of each form of bias exclusively focuses on the presence of its corresponding factor influences, excluding consideration of any other factors or biases.
Importantly, our definitions only consider what variables the propensities of ratings depend on.
Thereby, our usage of the terms is only concerned with the specific pattern the selection bias follows, and not with its resulting effects.
In this regard, our approach contrasts with prior work that identifies types of selection bias by the highly-skewed rating distributions that they can produce~\cite{pradel2012ranking,chen2023bias,steck2011item,abdollahpouri2019popularity}.
For example, a long-tailed rating distribution where a few items receive the most ratings (\eg Fig.~\ref{fig:yahoo:itempopularity}) is sometimes referred to as popularity bias or evidence thereof~\cite{chen2023bias,steck2011item,abdollahpouri2019popularity}.
Similarly, a difference between rating value frequencies from natural user behavior and ratings on randomly sampled items (\eg Fig.~\ref{fig:yahoo:rating}) is sometimes referred to as (evidence of) positivity bias~\cite{pradel2012ranking}.

However, these skewed distributions can occur for many reasons, and therefore, it is difficult to use their observation as evidence for a specific form of selection bias.
For example, a long-tailed rating distribution could result from positivity bias per Definition~\ref{def:pos-bias}: if there are only a few items with high rating values then these items will get the most ratings.
Vice versa, the differences between rating distributions could result from popularity bias per Definition~\ref{def:pop-bias} in a case where the more popular items happen to have a higher rating on average (see Fig.~\ref{fig:correlated_rating_item}).
To avoid this ambiguity and since our focus is on how selection bias should be modeled, we explicitly choose to base our definitions around the dependencies of propensities and will use the terms popularity bias and positivity bias accordingly.

\subsection{Rating prediction from user ratings}
Our goal is to optimize an \ac{RS} model that best predicts the user ratings across all items.
This is achieved by minimizing a loss function that compares the actual ratings $y_{u,i}$ and the predicted ratings $\hat{y}_{u,i}$:
\begin{equation}
    \mathcal{L} = \frac{1}{|\mathcal{U}| \, |\mathcal{I}|} \sum_{u\in \mathcal{U}} \sum_{i \in \mathcal{I}} \delta(\hat{y}_{u,i}, {y}_{u,i}),
    \label{eq:standard_loss}
\end{equation}
where the comparison function $\delta$ can be an \ac{RS} metric, \ie the com\-monly-used \ac{MSE}: $\delta(\hat{y}_{u,i}, y_{u,i}) = (\hat{y}_{u,i}- y_{u,i})^2$.

The loss function in Eq.~\ref{eq:standard_loss} represents our ideal goal but assumes that all ratings are available, something that is rarely the case in practice.
A straightforward but naive estimate of the ideal goal is to average over the observed ratings in the logged data $\mathcal{D}$:
\begin{equation}
    \mathcal{L}_\text{Naive} = \frac{1}{|\mathcal{D}|} \sum_{u,i \in \mathcal{D}} \delta(\hat{y}_{u,i}, y_{u,i}).
    \label{eq:naive_loss}
\end{equation}
However, this naive estimate ignores the effect of selection bias and assumes that every rating is equally probable to be observed~\cite{schnabel2016recommendations}.
As a result, 
if logged data $\mathcal{D}$ is subject to selection bias, it is biased by rating propensities:
\begin{equation}
        \mathbb{E}_o[\mathcal{L}_{\text{Naive}}] = \frac{1}{|\mathcal{D}|}  \sum_{u\in \mathcal{U}} \sum_{i \in \mathcal{I}} p_{u,i}\, \delta(\hat{y}_{u,i}, y_{u,i}) \not\propto \mathcal{L}.
\end{equation}

\subsection{IPS-based debiasing method}
To mitigate the effect of selection bias, widely-used methods make use of \acf{IPS}~\cite{imbens2015causal} and integrate it into the learning process~\cite{schnabel2016recommendations, huang2020keeping, joachims2017unbiased}. 
\ac{IPS} weights each rating inversely to its propensity, $p_{u,i}$, and, thereby, corrects for the over- and under-representation resulting from selection bias:
\begin{equation}
    \label{eq:ips}
    \mathcal{L}_\text{IPS} = \frac{1}{|\mathcal{U}| \,|\mathcal{I}|} \sum_{u,i \in \mathcal{D}} \frac{\delta(\hat{y}_{u,i}, y_{u,i})}{p_{u,i}}.
\end{equation}
Thus, \ac{IPS} gives more weight to observed ratings with small propensities and less weight to those with large propensities.
As $\mathbb{E}_o[o_{u,i}] = p_{u,i}$, the \ac{IPS} loss provides an unbiased estimate of $\mathcal{L}$:
\begin{equation}
    \mathbb{E}_o [\mathcal{L}_\text{IPS}] = \frac{1}{| \mathcal{U}| \,| \mathcal{I}|} \sum_u \sum_i \frac{\mathbb{E}_o[o_{u,i}]}{p_{u,i}} \delta(\hat{y}_{u,i}, y_{u,i}) = \mathcal{L}.
\end{equation}
Combined with a recommendation method, \eg \ac{MF}, \ac{IPS} reduces the effect of bias in predicting user ratings.

\subsection{Existing single-factor propensity estimation} %
\label{sec:single-factor-bias-estimations}
\ac{IPS} for rating estimation requires propensity estimation because propensities cannot be observed directly, since the exact way users decide to rate items is not directly accessible.
Methods exist for estimating propensities under our definitions of positivity bias (Definition~\ref{def:pos-bias}) and popularity bias (Definition~\ref{def:pop-bias}).
Importantly, each existing method only corresponds to one of the definitions and thus assumes that propensities only depend on a single factor.

The predominant method of positivity bias estimation in previous work 
uses Bayes' rule~\cite{schnabel2016recommendations}:
\begin{equation}
    \hat{p}_{u,i}^\text{pos} = P(o =1 \mid y = y_{u,i})
    = \frac{P(y = y_{u,i} \mid o =1) \, P(o =1)}{P(y = y_{u,i})}
    .
    \label{eq:pospropbayes}
\end{equation}
The observation prior is estimated by the observation frequency: $P(o =1) \approx |\mathcal{D}| / (|\mathcal{U}| \, |\mathcal{I}|)$, and the conditional rating-value probability estimate is the frequency of the rating in the observed data:  $P(y = r \mid o =1) \approx \sum_{u,i \in\mathcal{D}} \mathds{1}[y_{u,i}=r] / |\mathcal{D}|$.
Finally, to estimate the rating-value prior, a small sample of unbiased (\ac{MCAR}) data $\mathcal{M}$ is used; such data could be obtained by having users rate randomly sampled items. The prior estimate is simply the rating-value frequency in $\mathcal{M}$: $P(y = r) = \sum_{u,i \in \mathcal{M}} \mathds{1}[y_{u,i}=r]/ |\mathcal{M}|$.
Putting these components into Eq.~\ref{eq:pospropbayes}, we see that positivity bias propensities are estimated as follows:
\begin{equation}
    \hat{p}_{u,i}^\text{pos} \! = P(o =1 \,|\, y = y_{u,i}) \! \approx \! \frac{  |\mathcal{M}| \sum_{u',i' \in\mathcal{D}} \mathds{1} [ y_{u',i'}= y_{u,i} ] }{ |\mathcal{U}|  |\mathcal{I}| \sum_{u',i' \in \mathcal{M}} \mathds{1} [ y_{u',i'}= y_{u,i}] }
    .
\end{equation}

\noindent%
The most widely-used model of popularity bias
computes propensities on items based on item popularity~\cite{yang2018unbiased,saito2020unbiased}:
\begin{equation}
    \hat{p}_{u,i}^\text{pop} = P(o =1 \mid i ) \approx \frac{\sum_{u'} o_{u',i}}{\sum_{u'} \sum_{i'} o_{u',i'}}.
\end{equation}
These estimated propensities may be small, especially for tail items, thus causing high variance in the \ac{IPS} estimation.
Propensity clipping is usually used as a variance reduction technique~\citep{strehl2010learning}; it clips propensity scores by a small value $\tau$: $\bar{p}_{u,i} = \max(\hat{p}_{u,i}, \tau)$.
Here, $\tau$ trades off the bias and variance of the IPS estimation with the clipped estimated propensities:
If $\tau=1$, it approaches the naive estimation, while if $\tau=0$, it approaches the unbiased estimation.

With the corresponding estimated propensities, the \ac{IPS} estimator can be used to mitigate the effect of popularity bias or positivity bias.
However, existing single-factor forms of bias do not account for the fact that real-world user decisions toward rating items generally depend on more than one factor~\cite{pradel2012ranking,idrissiguess,ekstrand2018all}.

%% file: sections/sec-model.tex
\section{Correction for Multifactorial Bias}
\label{sec:our_method}
In contrast with existing single-factor models of bias, we consider a multifactorial bias that is determined by two factors: the item and rating value.
After defining our multifactorial bias, we introduce a stable propensity estimation method for it by adopting propensity smoothing technique. 
We use \ac{IPS}-based optimization with our novel estimated propensities, resulting in an unbiased rating prediction method that corrects for multifactorial bias.

\subsection{Definition of multifactorial bias}
Multifactorial bias occurs if the process that decides whether a user provides a rating is not a random selection and is determined by multiple factors.
In this paper, we consider a specific multifactorial bias that is determined by two factors: the item and rating value.
\begin{definition}[Multifactorial bias]
    \label{def:mul-bias}
    Logged rating data $\mathcal{D}$ is subject to \emph{multifactorial bias} if the propensities of ratings depend on which item they correspond to and their rating values (Fig.~\ref{fig:sub:mul-bias}):
    \begin{align}
            & \textit{Multifactorial-bias}(\mathcal{D}) \Longleftrightarrow \big( \textit{Selection-bias}(\mathcal{D}) \, \land  \\
            &\quad\;\; \forall u, u' \in\mathcal{U}, \forall i, i' \in \mathcal{I}, \; ( i=i'  \land y_{u,i} = y_{u', i'}) \longrightarrow p_{u,i} = p_{u',i'} \big). \nonumber
    \end{align}
\end{definition}

\noindent
This definition encompasses any selection bias determined by both item and rating value factors and can naturally be extended to various types of multifactorial bias.

\subsection{Propensity estimate for multifactorial bias}
\label{sec:mul-bias-estimation}
A novel method is required 
to estimate multifactorial propensities $p_{u,i} = P(o=1 \mid y=y_{u,i}, i)$ that vary over different combinations of items and rating values.
We propose to decompose the multifactorial propensity with Bayes' rule:
\begin{equation}
    \label{eq:mul:naive_bayes}
    \hat{p}_{u,i}^{\text{mul}} = P(o=1 \mid y=y_{u,i},i) = \frac{P(y=y_{u,i}, i \mid o=1) P(o=1)}{P(y=y_{u,i}, i)},
\end{equation}
and use a maximum likelihood estimate for each component.
Our observation prior estimate is the observation frequency: $P(o=1)\approx |\mathcal{D}| / (|\mathcal{U}| \, |\mathcal{I}|)$. %
Our conditional joint rating-value and item probability estimate is their frequency in the observation data $\mathcal{D}$:
\begin{equation}
    P(y=r, i \mid o=1) \approx \sum\nolimits_{u,i'\in \mathcal{D}} \mathbbm{1}[{i'=i \land y_{u,i'}=r}] / {|\mathcal{D}|},
\end{equation}
 and our joint rating-value and item prior estimate is their joint frequency in the small unbiased (\ac{MCAR}) data $\mathcal{M}$:
\begin{equation}
    P(y=r, i) \approx \sum\nolimits_{u,i'\in \mathcal{M}} \mathbbm{1}[{i'=i \land y_{u,i'}=r}] / {|\mathcal{M}|}.
\end{equation}
While conceptually this propensity estimation is straightforward, it brings a severe practical challenge as it relies on the frequencies of combinations of items and rating values in the sparse observation data $\mathcal{D}$ and the even sparser unbiased data $\mathcal{M}$.
As a result, estimates of the joint probabilities can be extremely small or even zero, and, thereby, potentially result in invalid propensity estimates or extremely-high-variance \ac{IPS} estimates.

To address these sparsity issues, we apply Laplace smoothing~\cite{christopher2008introduction} to both the estimations of the joint conditional probability and joint prior.
The conditional joint rating-value and item probability estimate is smoothed with parameter $\alpha_1$:
\begin{equation}
    P(y=r, i \mid o=1) \approx \frac{\sum_{u,i'\in \mathcal{D}} \mathbbm{1}[{i'=i \land y_{u,i'}=r}] + \alpha_1}{|\mathcal{D}| + \alpha_1 |\mathcal{I}|\, |\mathcal{R}|}. \label{eq:smooth1} 
\end{equation}
The estimated joint rating-value and item prior is smoothed by $\alpha_2$:
\begin{align}
        & P(y=r, i) \approx \nonumber \\
        & \underbrace{\frac{\sum_{u,i'\in \mathcal{M}} \mathbbm{1}[{y_{u,i'}=r}]}{ |\mathcal{M}|}}_{\text{Estimate of } P(y=r).}
        \cdot \underbrace{\frac{\sum_{u,i'\in \mathcal{M}} \mathbbm{1}[i'=i \land y_{u,i'}=r] + \alpha_2}{\sum_{u,i'\in \mathcal{M}} \mathbbm{1}{[y_{u,i'}=r]}+\alpha_2 |\mathcal{I}|}}_{\text{Smoothed estimate of } P(i \mid y=r).} .
    \label{eq:smooth2}
\end{align}
Instead of directly smoothing the joint prior $P(y=r, i)$, we decompose it into the product of the prior $P(y=r)$ and the conditional $P(i\mid y=r)$ and only smooth the latter.
We found that this provided the most robust performance; most likely because item sparsity is much more extreme than rating-value sparsity.

\subsection{A debiasing method for multifactorial bias}
Using the results of our multifactorial bias propensity estimation, a rating prediction model can be optimized with \ac{IPS} while accounting for multifactorial bias.
Following~\citet{schnabel2016recommendations}, we choose inverse-propensity-scored \acl{MF} (MF-IPS) as the de-biased rating prediction method.
With the propensity estimates $\hat{p}_{u,i}^\text{mul}$, we have our multifactorial method: $\text{MF-IPS}^{Mul}$.
It minimizes the multifactorial \ac{IPS} estimate of the \ac{MSE} between the predicted ratings and the actual ratings with an added $L_2$-regularization term:
\begin{equation}
    \mathcal{L}_{\text{MF-IPS}^{Mul}} (\Theta) = \frac{1}{|\mathcal{D}|} \sum_{u,i\in\mathcal{D}}\frac{\delta(\hat{y}_{u,i}, y_{u,i})}{\hat{p}_{u,i}^\text{mul}} + \lambda||\Theta||^2_2,
\end{equation}
where a predicted rating is computed by a standard \ac{MF}: $\hat{y}_{u,i} = \bm{p}_u^\top \bm{q}_{i} + a_u + b_i + c$, 
which is the inner-product of embedding vectors $\bm{p}_u$ and $\bm{q}_{i}$ for user $u$ and item $i$, together with user, item and global offsets $a_u, b_i$ and $c$;
and the parameter set $\Theta=\{\bm{p}_u, \bm{q}_{i}, a_u, b_i, c\}$ includes all parameters of \ac{MF}.

In the optimization of our multifactorial method, we could follow common stochastic gradient descent and iteratively sample a batch of data and update parameter $\theta \in \Theta$ according to gradient of the loss function on each data batch using the Adam optimizer~\cite{Kingma2015Adam}:
\begin{equation}
    \label{eq:gradient}
    \theta_{t} = \text{ADAM} (\theta_{t-1}, \nabla_{\theta_{t-1}} \mathcal{L}_{\text{MF-IPS}^{Mul}}).
\end{equation}
However, we found this concurrent gradient descent method in \ac{IPS}-based optimization to be unstable in experiments on real-world data (see Section~\ref{sec:experiments:real_world}).
Many data batches contain widely varied propensity estimates,
and due to the very low propensities under multifactorial bias, this appears to result in severe instability between updates.

\begin{algorithm}[t]
    \DontPrintSemicolon
    \KwIn{Observed rating data: $\mathcal{D}$; estimated propensities: $\hat{p}$.}
    \KwOut{$\text{MF-IPS}^{Mul}$ parameters: $\bm{p}_u, \bm{q}_i, a_u, b_i, c$.}
    Initialize parameters $\bm{p}_u, \bm{q}_i, a_u, b_i, c$; \;
    \While{stop condition is not reached}{
      \tcc{Epoch to update global \& user embeddings and offsets.}
      \For{each batch of $(u,i,y_{u,i})$ in a random ordering of $\mathcal{D}$}
      {\label{line:opt_start:user-emb}
        Update parameters $\bm{p}_u, a_u, c$ according to Eq.~\ref{eq:gradient};
      }
      \label{line:opt_end:user-emb}
      \tcc{Epoch to update item embeddings and offsets.}
      \For{each batch of $(u,i,y_{u,i})$ in a random ordering of $\mathcal{D}$}
      {\label{line:opt_start:item-emb}
        Update parameters $\bm{q}_i, b_i$ according to Eq.~\ref{eq:gradient};
      }
      \label{line:opt_end:item-emb}
    }
    \label{line:epoch:end}
    \caption{Our optimization method for $\text{MF-IPS}^{Mul}$ with our alternating gradient descent approach.}\label{alg:ALS}
\end{algorithm}

An existing alternative to the concurrent gradient descent is the \acf{ALS} method~\cite{takacs2012alternating}.
\ac{ALS} iteratively alternates between optimizing user and item embeddings via least squares to reduce optimization instability.
The alternating updates mitigate the effect of noise and outlier interactions~\cite{takacs2012alternating}.
We build on the idea of alternating gradient descent from \ac{ALS} and extend it to optimize generic loss functions using the Adam optimizer.
Algorithm~\ref{alg:ALS} shows the procedure of optimizing $\text{MF-IPS}^{Mul}$ with our alternating gradient descent method.
It begins with parameter initialization, then updates parameters over multiple epochs according to the loss on logged user ratings $\mathcal{D}$.
The optimization continues until the stop condition is reached, \eg decreasing performance on the validation set or reaching a predefined number of epochs.
Importantly, in each epoch, the item-related parameters $\bm{q}_i, b_i$ (line~\ref{line:opt_start:item-emb}--\ref{line:opt_end:item-emb}) and other parameters $\bm{p}_u, a_u, c$ (line~\ref{line:opt_start:user-emb}--\ref{line:opt_end:user-emb}) are updated independently and alternately.
Thereby, our optimization alternately updates a subset of parameters while keeping the remaining parameters fixed in each epoch.
Our experimental results on real-world data indicate this leads to increased stability and robustness (see Section~\ref{sec:experiments:real_world}).

This completes the description of our method to mitigate the effects of multifactorial bias.
It optimizes a \ac{MF} model for rating predictions using \ac{IPS} with multifactorial bias propensity estimation that considers both item and rating value factors.
In addition, we adopt propensity smoothing and alternating gradient descent to make our multifactorial method feasible and robust in practice.

%% file: sections/sec-experiment.tex
\section{Experiments on Real-world Data}
\label{sec:experiments:real_world}
Our experimental analysis on real-world datasets aims to answer two research questions:
\begin{enumerate*}[label=(\textbf{RQ\arabic*})] 
    \item Does our proposed multifactorial method better mitigate the effect of bias in logged rating data than existing single-factor debiasing methods? %
    \item How do varying smoothing parameters and our alternating gradient descent approach affect our multifactorial method?
\end{enumerate*}

\subsection{Experimental setup}

\begin{table*}[ht]
	\centering
        \caption{Performance comparison for predicting ratings on the Yahoo!R3 and Coat datasets. 
        Results are means of 10 independent runs with standard deviations in brackets.
        $\dag$ indicates that our multifactorial method $\text{MF-IPS}^{Mul}$ with alternating gradient descent significantly outperforms all other existing methods (paired-samples t-test $(p < 0.01)$).
        } %
		\label{tab:overall}%
        \begin{subfigure}{\linewidth}
            \centering
			\begin{tabular}{l l l@{~}l l@{~}l l@{~}l l@{~}l l@{~}l}
			\toprule
			Dataset & Method & \multicolumn{2}{l}{MSE} & \multicolumn{2}{l}{MAE} & \multicolumn{2}{l}{RMSE} & \multicolumn{2}{l}{$\text{RMSE}_U$} & \multicolumn{2}{l}{$\text{RMSE}_I$}\\
            \midrule
            \multirow{8}{*}{{\emph{Yahoo!R3}}} 
            & Avg & 2.1321 & & 1.2671 & & 1.4602 & & 1.4167 & & 1.4153  \\ 
            & MF & 1.8296 & \footnotesize (0.0318) & 1.1305 & \footnotesize (0.0173) & 1.3526 & \footnotesize (0.0117) & 1.2593 & \footnotesize (0.0159) & 1.3325 & \footnotesize (0.0130) \\ 
            & VAE & 1.4182 & \footnotesize (0.0082) & 0.9677 & \footnotesize (0.0039) & 1.1909 & \footnotesize (0.0034) & 1.1158 & \footnotesize (0.0034) & 1.1694 & \footnotesize (0.0033) \\
            \cmidrule{2-12}
            & $\text{MF-IPS}^{MF}$ & 1.7877 & \footnotesize (0.0297) & 1.0621 & \footnotesize (0.0024) & 1.3370 & \footnotesize (0.0111) & 1.2140 & \footnotesize (0.0050) & 1.3067 & \footnotesize (0.0109) \\         
            & $\text{MF-IPS}^{Pop}$ & 1.9432 & \footnotesize (0.0048) & 1.1425 & \footnotesize (0.0058) & 1.3940 & \footnotesize (0.0017) & 1.2783 & \footnotesize (0.0046) & 1.3711 & \footnotesize (0.0008) \\
            & $\text{MF-IPS}^{Pos}$ & 0.9891 & \footnotesize (0.0013) & 0.7928 & \footnotesize (0.0079) & 0.9945 & \footnotesize (0.0006) & 0.9267 & \footnotesize (0.0048) & 0.9774 & \footnotesize (0.0015) \\
            \cmidrule{2-12}
            & $\text{MF-IPS}^{Mul}$ (ours) & \textbf{0.9629}$^\dag$ & \footnotesize (0.0015) & \textbf{0.7700}$^\dag$ & \footnotesize (0.0120) & \textbf{0.9813}$^\dag$ & \footnotesize (0.0007) & \textbf{0.9071}$^\dag$ & \footnotesize (0.0075) & \textbf{0.9626}$^\dag$ & \footnotesize (0.0025) \\
            \midrule
            \multirow{8}{*}{{\emph{Coat}}} 
            & Avg & 1.6521 & & 1.0904 & & 1.2854 & & 1.2521 & & 1.2605  \\ 
            & MF & 1.2916 & \footnotesize (0.0108) & 0.9283 & \footnotesize (0.0074) & 1.1365 & \footnotesize (0.0048) & 1.0907 & \footnotesize (0.0049) & 1.1085 & \footnotesize (0.0049) \\ 
            & VAE & 1.1393 & \footnotesize (0.0048) & 0.8583 & \footnotesize (0.0038) & 1.0674 & \footnotesize (0.0023) & 1.0282 & \footnotesize (0.0027) & 1.0424 & \footnotesize (0.0021) \\
            \cmidrule{2-12}
            & $\text{MF-IPS}^{MF}$ & 1.1597 & \footnotesize (0.0175) & 0.8687 & \footnotesize (0.0165) & 1.0769 & \footnotesize (0.0082) & 1.0366 & \footnotesize (0.0076) & 1.0512 & \footnotesize (0.0074) \\
            & $\text{MF-IPS}^{Pop}$ & 1.2284 & \footnotesize (0.0142) & 0.9042 & \footnotesize (0.0115) & 1.1083 & \footnotesize (0.0064) & 1.0666 & \footnotesize (0.0066) & 1.0828 & \footnotesize (0.0066) \\
            & $\text{MF-IPS}^{Pos}$ & 1.1728 & \footnotesize (0.0120) & 0.8708 & \footnotesize (0.0129) & 1.0830 & \footnotesize (0.0055) & 1.0395 & \footnotesize (0.0073) & 1.0576 & \footnotesize (0.0069) \\
            \cmidrule{2-12}
            & $\text{MF-IPS}^{Mul}$ (ours) & \textbf{1.1020}$^\dag$ & \footnotesize (0.0007) & \textbf{0.8552}$^\dag$ & \footnotesize (0.0023) & \textbf{1.0498}$^\dag$ & \footnotesize (0.0003) & \textbf{1.0110}$^\dag$ & \footnotesize (0.0009) & \textbf{1.0275}$^\dag$ & \footnotesize (0.0006) \\
              \bottomrule
           \end{tabular}
    \end{subfigure}
    \vspace{-10pt}
\end{table*}%

Our experiments are based on two real-world datasets: \text{Yahoo!R3}~\cite{marlin2009collaborative} and Coat~\cite{schnabel2016recommendations}, which are publicly available and widely used to evaluate debiasing methods.\footnote{The KuaiRec dataset~\cite{gao2022kuairec} contains biased user interactions (a sparse subset) and a set of fully observed user-item interactions (a dense subset). However, as highlighted by \citet{lin2023transfer}, its density (16.3\% and 99.9\% for the sparse and dense subsets, respectively) surpasses that of other datasets, diverging from our targeted bias and sparsity problem, as explained in Section~\ref{section:introduction}. We extend our evaluation beyond the Yahoo!R3 and Coat datasets by conducting a \emph{simulation} on KuaiRec in Section~\ref{sec:simulation}.}
Both have a training set consisting of biased ratings and a \ac{MCAR} test set of user ratings on uniformly randomly selected items.
We filter the users that do not appear in the test sets to make predictions more precise, resulting in 129,179 biased ratings and 54,000 unbiased ratings of 5,400 users to 1,000 items in the \text{Yahoo!R3} dataset, 
and 6,960 biased ratings and 4,640 unbiased ratings of 290 users to 300 items in the Coat dataset, respectively.
The biased ratings are partitioned into a training and validation set according to a ratio of \text{4:1}.
To estimate propensities, we set aside 5\% and 20\% of the original test sets as the small unbiased data $\mathcal{M}$ for the \text{Yahoo!R3} and Coat datasets, respectively.
This ensures at least two interactions per item for estimating the conditional joint rating-value and item distribution.

To evaluate our method, we adopt evaluation metrics widely used in previous work~\cite{schnabel2016recommendations,wang2019doubly,steck2013evaluation}: \ac{MSE}, \ac{RMSE}, and \ac{MAE}.
We further report the average RMSE performance per user ($\text{RMSE}_U$) and item ($\text{RMSE}_I$)~\citep{massa2007trust}, \ie we calculate the RMSE score for each individual user/item separately and then average them.

We evaluate our multifactorial method $\text{MF-IPS}^{Mul}$ by comparing it with the following baselines: 
\begin{enumerate*}[label=(\roman*), nosep, leftmargin=*]
	\item \acused{Avg}\ac{Avg}, MF, and VAE~\cite{liang2018variational} that ignore bias altogether. 
    \ac{Avg} simply predicts the average observed rating of each item: $\hat{y}_{u,i} = \frac{\sum_{u',i \in \mathcal{D}} y_{u',i}}{|\{(u',i,y_{u',i})\in \mathcal{D}\}|}$. VAE has been proposed to apply variational autoencoders to collaborative filtering. We adopt Gaussian log-likelihood in VAE for rating predictions.
	\item $\text{MF-IPS}^{MF}$, a debiased method with propensity estimation using \ac{MF} with logistic regression~\citep{saito2020unbiased,huang2022different,schnabel2016recommendations}. 
	It has the potential to correct for multifactorial bias as it uses \ac{MF} to model bias through learned multiple hidden factors.
	\item $\text{MF-IPS}^{Pop}$ and $\text{MF-IPS}^{Pos}$, two debiased methods with single-factor popularity bias estimation and single-factor positivity bias estimation, respectively.
\end{enumerate*}	

Additionally, to evaluate how our proposed alternating gradient descent approach affects rating prediction models,
all \ac{MF}-based models are optimized by two optimization methods: 
\begin{enumerate*}[label=(\roman*)]
	\item \emph{Concurrent} gradient descent: all parameters of methods are updated concurrently;
	\item \emph{Alternating} gradient descent: the item-related parameters and other parameters are updated alternately.
\end{enumerate*}

Hyperparameters used in the \ac{MF}-based methods are tuned per propensity estimation in the following range:
the learning rate $\eta \in \{10^{-3}, 10^{-4}, 10^{-5}\}$, the $L_2$ regularization weights $\lambda \in \{10^{-7}, 10^{-6},$ $\ldots, 10^{-2}\}$, and the dimension of embeddings of users and items $d\in \{16, 32, 64, 128\}$.
Hyperparameter tuning for VAE is conducted as follows: 
(i) aligning the learning rate, the regularization weights, and the dimension of the latent representation with the same range employed for \ac{MF}-based methods; 
and (ii) adjusting the parameter that controls the strength of the Kullback-Leibler term within the range $\{0.05, 0.1, 0.2, 0.4, \ldots, 1.0\}$.
For debiasing methods with multifactorial bias estimation, we also choose the smoothing parameters $\alpha_1, \alpha_2 \in \{1, 2, \ldots, 10\}$.
Additionally, propensity clipping and normalization are used to reduce variance and improve the robustness of methods.
Our experimental implementation 
is available at \url{https://github.com/BetsyHJ/MultifactorialBias}.

\begin{table*}[t]
	\centering
        \caption{Performance comparison among \ac{MF}-based methods when optimization is done with concurrent and alternating gradient descent on the Yahoo!R3 and Coat datasets. 
        Results are means of 10 independent runs with standard deviations in brackets.
        $\dag$ indicates that the method optimized by the alternating gradient descent method significantly outperforms the identical method optimized by the concurrent gradient descent method (paired-samples t-test $(p < 0.01)$).} %
		\label{tab:alternating}%
        \begin{subfigure}{\linewidth}
            \centering
			\begin{tabular}{l l l@{~}l l@{~}l l@{~}l l@{~}l l@{~}l l@{~}l}
			\toprule
            \multirow{2}[1]{*}{Dataset} & \multirow{2}[1]{*}{Method} & \multicolumn{6}{c}{Concurrent} & \multicolumn{6}{c}{Alternating} \\
            \cmidrule(r){3-8} \cmidrule(r){9-14}
			 & & \multicolumn{2}{l}{MSE} & \multicolumn{2}{l}{MAE} & \multicolumn{2}{l}{RMSE} &\multicolumn{2}{l}{MSE} & \multicolumn{2}{l}{MAE} & \multicolumn{2}{l}{RMSE} \\
            \midrule
            \multirow{6}{*}{\emph{Yahoo!R3}}
            & MF & 1.8296 & \footnotesize (0.0318) & 1.1305 & \footnotesize (0.0173) & 1.3526 & \footnotesize (0.0117) & 1.8335 & \footnotesize (0.0236) & 1.1688 & \footnotesize (0.0077) & 1.3540 & \footnotesize (0.0088) \\ 
            \cmidrule{2-14}
            & $\text{MF-IPS}^{MF}$ & 1.7877 & \footnotesize (0.0297) & 1.0621 & \footnotesize (0.0024) & 1.3370 & \footnotesize (0.0111) & 1.7143$^\dag$ & \footnotesize (0.0172) & 1.0616 & \footnotesize (0.0168) & 1.3093$^\dag$ & \footnotesize (0.0066) \\         
            & $\text{MF-IPS}^{Pop}$ & 1.9432 & \footnotesize (0.0048) & 1.1425 & \footnotesize (0.0058) & 1.3940 & \footnotesize (0.0017) & 1.9055$^\dag$ & \footnotesize (0.0196) & 1.1659 & \footnotesize (0.0077) & 1.3804$^\dag$ & \footnotesize (0.0071) \\
            & $\text{MF-IPS}^{Pos}$ & 0.9891 & \footnotesize (0.0013) & 0.7928 & \footnotesize (0.0079) & 0.9945 & \footnotesize (0.0006) & {0.9762}$^\dag$ & \footnotesize (0.0034) & 0.7943 & \footnotesize (0.0099) & {0.9880}$^\dag$ & \footnotesize (0.0017) \\
            \cmidrule{2-14}
            & $\text{MF-IPS}^{Mul}$ (ours) & 0.9812 & \footnotesize (0.0067) & {0.7737} & \footnotesize (0.0116) & 0.9905 & \footnotesize (0.0034) & \textbf{0.9629}$^\dag$ & \footnotesize (0.0015) & \textbf{0.7700} & \footnotesize (0.0120) & \textbf{0.9813}$^\dag$  & \footnotesize (0.0007) \\
            \midrule
            \multirow{6}{*}{\emph{Coat}}
            & MF & 1.2916 & \footnotesize (0.0108) & 0.9283 & \footnotesize (0.0074) & 1.1365 & \footnotesize (0.0048) & 1.2040$^\dag$ & \footnotesize (0.0119) & 0.9034$^\dag$ & \footnotesize (0.0208) & 1.0973$^\dag$ & \footnotesize (0.0054) \\ 
            \cmidrule{2-14}
            & $\text{MF-IPS}^{MF}$ & 1.1597 & \footnotesize (0.0175) & 0.8687 & \footnotesize (0.0165) & 1.0769 & \footnotesize (0.0082) & 1.1641 & \footnotesize (0.0154) & 0.8730 & \footnotesize (0.0287) & 1.0789 & \footnotesize (0.0072) \\
            & $\text{MF-IPS}^{Pop}$ & 1.2284 & \footnotesize (0.0142) & 0.9042 & \footnotesize (0.0115) & 1.1083 & \footnotesize (0.0064) & 1.1923$^\dag$ & \footnotesize (0.0049) & 0.8787$^\dag$ & \footnotesize (0.0124) & 1.0919$^\dag$ & \footnotesize (0.0022) \\
            & $\text{MF-IPS}^{Pos}$ & 1.1728 & \footnotesize (0.0120) & 0.8708 & \footnotesize (0.0129) & 1.0830 & \footnotesize (0.0055) & 1.1717 & \footnotesize (0.0065) & 0.8672  & \footnotesize (0.0106) & 1.0825 & \footnotesize (0.0030) \\
            \cmidrule{2-14}
            & $\text{MF-IPS}^{Mul}$ (ours) & {1.1397} & \footnotesize (0.0295) & \textbf{0.8503} & \footnotesize (0.0199) & {1.0675} & \footnotesize (0.0138) & \textbf{1.1020}$^\dag$ & \footnotesize (0.0007) & {0.8552} & \footnotesize (0.0023) & \textbf{1.0498}$^\dag$ & \footnotesize (0.0003) \\ 
              \bottomrule
           \end{tabular}
    \end{subfigure}
    \vspace{-10pt}
\end{table*}%

\subsection{Overall performance}
Table~\ref{tab:overall} displays our main experimental results on the \text{Yahoo!R3} and Coat datasets.
We make the following three observations.
    First, among all the methods, \ac{Avg} has the worst performance; this is expected as it provides non-personalized predictions and ignores selection bias.
    Accordingly, \ac{MF} does model individual user preferences and outperforms \ac{Avg}.
    Importantly, VAE exhibits a considerable performance margin over \ac{MF} due to its generative capabilities.

    Second, the debiasing methods that consider the effect of bias improve the performance: MF-IPS $\succ$ MF (except for $\text{MF-IPS}^{Pop}$ on \text{Yahoo!R3}).\footnote{We write $A\succ B$ to indicate that method $A$ outperforms method $B$.} 
    A strong indication of the negative effect that selection bias has on rating prediction optimization. %
    
    Third, in debiasing methods, positivity bias estimation performs better than popularity bias estimation, but worse than multifactorial bias estimation: $\text{MF-IPS}^{Mul}$ $\succ$ $\text{MF-IPS}^{Pos}$ $\succ$ $\text{MF-IPS}^{Pop}$.
    This suggests that positivity bias has a stronger effect than popularity bias in rating predictions.
    Despite the potential to capture multifactorial forms of bias, $\text{MF-IPS}^{MF}$ does not always outperform $\text{MF-IPS}^{Pos}$, suggesting that it cannot adequately learn multifactorial bias.
    Nevertheless, multifactorial bias estimation provides the most robust and best overall performance; $\text{MF-IPS}^{Mul}$ significantly outperforms all other methods on both datasets.
	By considering the effect of multiple factors on selection bias, the multifactorial method can better capture and correct for bias in real-world data.

Overall, the best-performing method is our multifactorial debiasing method with alternating gradient descent. 
Therefore, we answer \textbf{RQ1} in the affirmative: 
The proposed multifactorial method $\text{MF-IPS}^{Mul}$ better mitigates the effect of bias in logged rating data than methods designed for single-factor biases.

\vspace*{-1mm}
\subsection{Smoothing and alternating gradient descent}
\begin{figure}[t]
    \centering
    \includegraphics[width=0.65\linewidth]{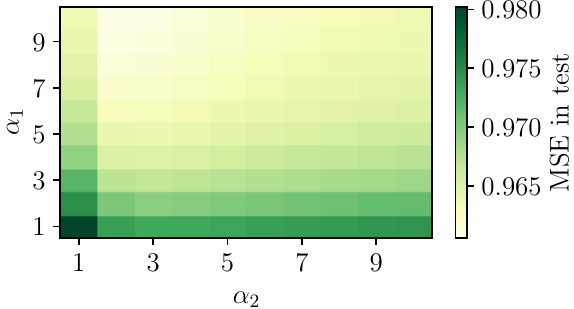}%
    \caption{(\text{Yahoo!R3}) The effect of varying smoothing parameters $\alpha_1$ and $\alpha_2$ on MSE obtained by our multifactorial method.}
    \label{fig:smoothing}
\end{figure}

To better understand the effect of propensity smoothing and alternating gradient descent, we perform the following additional analyses.
Due to space limitations, some analyses are limited to the \text{Yahoo!R3} dataset only. 
First, we look at how the performance of our multifactorial method changes when varying the smoothing parameters.
Fig.~\ref{fig:smoothing} shows the \ac{MSE} performance obtained for different smoothing parameters: $\alpha_1$ and $\alpha_2$ (see Eq.~\ref{eq:smooth1} and Eq.~\ref{eq:smooth2}).
We see that the highest performance is reached with $\alpha_1 = 10$ and $\alpha_2 = 2$, however, there is clearly a wide range of smoothing parameters that provide close to optimal performance.
It appears that it is mainly important not to set the parameters too small, as the worst performance is reached with $\alpha_1 = 1$ and $\alpha_2 = 1$.
The combined results of Fig.~\ref{fig:smoothing} and Table~\ref{tab:overall} reveal that the smoothing parameters do not need fine-tuning for the multifactorial method to outperform all other methods.
Thus, we conclude that propensity smoothing is an effective and robust enhancement for multifactorial debiasing.

Second, we compare \ac{MF}-based methods optimized by the concurrent method against those optimized by the alternating method, as shown in Table~\ref{tab:alternating}.
These performance improvements are considerably enhanced with our alternating gradient descent method, which boosts the performance of $\text{MF-IPS}^{Mul}$ on all datasets and all metrics (with the exception of MAE on Coat).
Performance gains are also seen for other methods but not as consistent as for $\text{MF-IPS}^{Mul}$.
Due to the smaller multifactorial propensities, $\text{MF-IPS}^{Mul}$ has more variance during optimization, and therefore, alternating gradient descent can provide a more consistent improvement here.

\begin{figure}[t]
    \centering
    \begin{subfigure}{0.485\linewidth}
        \includegraphics[width=\linewidth]{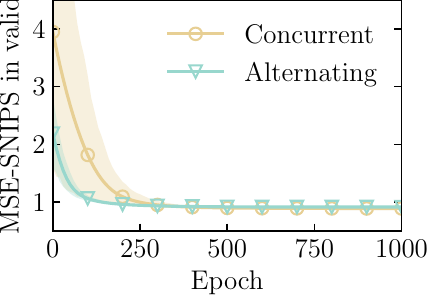}
    \label{fig:yahoo:lc_valid}
    \vspace{-7pt}
    \end{subfigure}
    \begin{subfigure}{0.505\linewidth}
        \includegraphics[width=\linewidth]{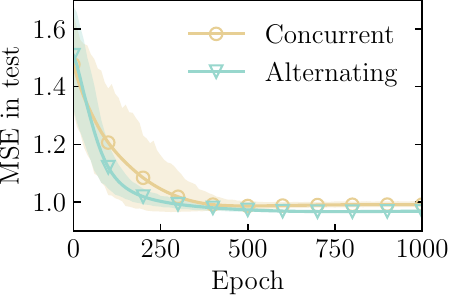}
    \label{fig:yahoo:lc_test}
    \vspace{-7pt}
    \end{subfigure}
    \caption{(Yahoo!R3)
    Learning curves tracking self-normal\-ized \ac{IPS}-weighted \ac{MSE} on the validation set and \ac{MSE} on the test set obtained by our multifactorial method. 
    Results are means over 10 independent runs, shared areas show the 95\% confident intervals calculated by using bootstrapping~\cite{diciccio1996bootstrap}.}
    \label{fig:yahoo:learning_curve}
\end{figure}

We further compare the learning curves of our multifactorial method when optimization is done with the concurrent and alternating gradient descent.
Fig.~\ref{fig:yahoo:learning_curve} displays these in terms of the self-normalized \ac{IPS}-weighted \ac{MSE} performance~\citep{swaminathan2015self} on the validation set and the \ac{MSE} performance on the test set.
Clearly, the alternating method exhibits more stable and faster learning than the concurrent method in the early stages of learning.
While both converge around 500 epochs,
the concurrent method converges to a slightly better MSE-IPS performance on the validation set compared to the alternating method.
However, we see that this actually results in a slightly worse \ac{MSE} performance on the test set, suggesting the concurrent method is more prone to overfitting.
Therefore, it appears that alternating gradient descent is indeed less influenced by noise and outliers than the concurrent method, which we think is why it provides more stable and robust optimization.

Finally, we answer \textbf{RQ2}:
propensity smoothing provides robust performance improvements to our multifactorial method and does not need fine-tuning;
alternating gradient descent leads to less variance in learning curves and less overfitting than concurrent gradient descent.
These advantages substantially increase the robustness, stability, and performance of our multifactorial method.

%% file: sections/sec-biases-effect.tex
\section{Effect of Biases on User Ratings}
\label{sec:simulation}
We turn to our final research question:
\begin{enumerate*}[label=(\textbf{RQ\arabic*})] 
\setcounter{enumi}{2}
 \item Can our multifactorial method $\text{MF-IPS}^{Mul}$ robustly mitigate the effect of selection bias in scenarios where the effect of two factors on bias is varied?
\end{enumerate*}

\vspace*{-1mm}
\subsection{Experimental setup for \textbf{RQ3}}
Due to a lack of real-world datasets with different effects of each factor, we utilize a semi-synthetic setup.
We simulate a short-video rating scenario by sampling user ratings on videos under different forms of selection bias.
Our sampling source is the KuaiRec dataset~\cite{gao2022kuairec} as it provides a fully observed user-item interaction matrix where 1,411 users rate almost all 3,327 items.

Since the dataset does not contain ratings but watch ratios on videos, we first convert these into 5-star user ratings.
First, we sort the watch ratios in ascending order and then give the top 51.48\% a rating of $y=1$, the next 25.25\% get $y=2$, etc.,
such that the resulting ratings follow the rating distribution of the {Yahoo!R3} dataset:
$P(y=1) = 0.5148$,
$P(y=2) = 0.2525$,
$P(y=3) = 0.1496$,
$P(y=4) = 0.0554$ and
$P(y=5) = 0.0277$.

The biased training set is constructed by sampling ratings with multifactorial selection bias.
To simulate the joint effect of rating value and item factors, we first introduce two single-factor propensities: $\rho^{(\text{R})}$ which is only dependent on the rating values, and $\rho^{(\text{I})}$ which is only dependent on the items.
Our simulated multifactorial propensity is then simply a linear interpolation between the two:
\begin{equation}
    P(o=1 \mid y=r, i) = \gamma \rho^{(\text{R})}_r + (1 - \gamma) \rho^{(\text{I})}_i, \label{eq:simulation}
\end{equation}
where $\gamma \in [0, 1]$ controls the effect of each factor on the selection bias.
Our simulation also covers single-factor scenarios:
if $\gamma=0.0$, the selection bias is \emph{popularity bias}, only determined by the item factor;
if $\gamma=1.0$, it is \emph{positivity bias}, only determined by the rating value factor.
Importantly, when $\gamma \in (0, 1)$, the resulting selection bias is multifactorial as it is affected by both factors.

Our rating-value propensities are
$\rho^{(\text{R})} = [0.0123, 0.0102, 0.0213,$ $0.0568, 0.1795]$ corresponding to the ratings $[1,2,3,4,5]$.
These values were chosen to match the positivity bias propensities estimated on the \text{Yahoo!R3} datasets, and they lead to an expectation of ratings higher than 3 being over-represented.
Item propensities are generated according to a power-law distribution following~\citet{bellogin2017statistical}:
$\rho^{(\text{I})} = (\eta - 1)\cdot (\text{rank}(i)/k_\text{min})^{-\eta}$, where $\text{rank}(i) \in [1, |\mathcal{I}|]$ is the position of item $i$ when sorted by their average ratings descendingly, and
we set the power-law exponent $\eta = 1.4$ and the minimum value $k_\text{min} = 20$.
Hereby, more popular items have a higher rating on average as is often seen in real-world data (\eg Fig.~\ref{fig:correlated_rating_item}).

Some of our methods need a small unbiased \ac{MCAR} set and we need an unbiased test set for evaluation.
We sample unbiased data by uniform-randomly selecting 40 ratings from each user's ratings across all items.
From this data, we set aside 20\% for the small \ac{MCAR} set and use the remaining 80\% as the test set.

To answer \textbf{RQ3}, we compare the performance of our multifactorial method $\text{MF-IPS}^{Mul}$ to that of \ac{MF} with and without debiasing methods for single-factor bias correction: $\text{MF-IPS}^{Pop}$ and $\text{MF-IPS}^{Pos}$.
Additionally, we also consider debiasing with the ground truth propensities: $\text{MF-IPS}^{GT}$.
This provides an unrealistic skyline that is only possible in a simulation setting where the true propensities are known.
Due to space limitations, 
we only report \ac{MSE} and \ac{MAE} under optimization with alternating gradient descent.

\vspace*{-1mm}
\subsection{Results for RQ3}

\begin{figure*}[t]
    \centering
    \includegraphics[clip, trim=0mm 5.8mm 0mm 0mm, width=1\linewidth]{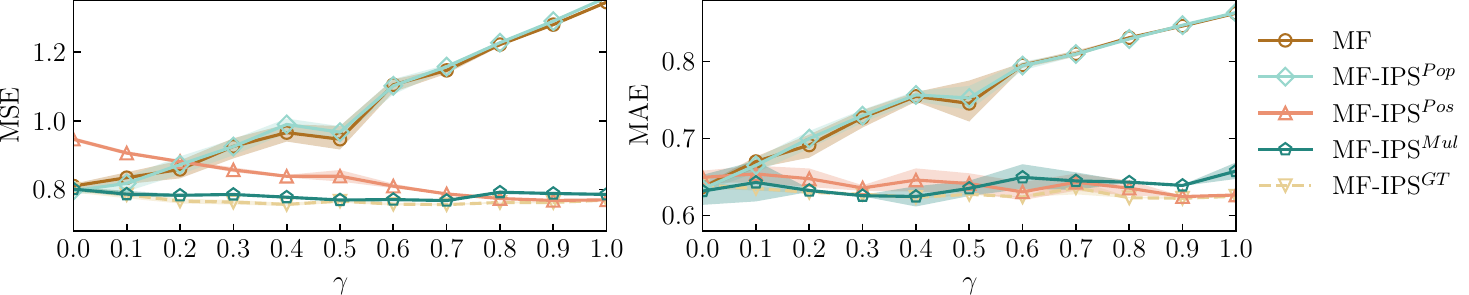}
    \caption{Performance in our simulated setting with different dependencies of bias on item and rating value factors through varying $\gamma$ (x-axis, Eq.~\ref{eq:simulation}). Results are means over 10 independent runs; shared areas show 95\% bootstrap confident intervals~\cite{diciccio1996bootstrap}.}
    \vspace{-13pt}
    \label{fig:sim3-results}
\end{figure*}

Fig.~\ref{fig:sim3-results} shows the performance of the different MF with various debiasing methods, under multifactorial selection bias, as $\gamma$ varies the effects of the rating-value and item factors.

We first consider when $\gamma$ equals 0, and the simulated selection bias reduces to popularity bias. 
Here, we see that $\text{MF-IPS}^{Pos}$ performs worst and that
      $\text{MF-IPS}^{Mul}$ and $\text{MF-IPS}^{Pop}$ have performance comparable and similar to MF.
This shows that assuming selection bias is dependent on only the rating value factor can substantially hurt performance when it is actually only dependent on the item factor.
However, it appears that assuming dependency on both factors does not hurt performance at all, in this scenario.

Next, we consider when $\gamma$ equals 1, and the simulated selection bias reduces to positivity bias. 
Here, we observe that
\ac{MF} and $\text{MF-IPS}^{Pop}$ perform worse than all other methods by a large margin;
and that
$\text{MF-IPS}^{Pos}$ has the best performance, while
our multifactorial method $\text{MF-IPS}^{Mul}$ performs slightly worse.
This strongly suggests that assuming selection bias is dependent only on the item factor is detrimental to performance when it is in fact dependent only on the rating value factor.
In contrast, the multifactorial model also made an incorrect assumption: a dependency on both factors, but this only resulted in a relatively small performance decrease.

Finally, we turn our attention to all other cases: where $\gamma \in (0, 1)$ and the selection bias is multifactorial bias.
We see that as $\gamma$ gets closer to $0$ or $1$, the performance of the corresponding single-factor debiasing method increases.
In contrast, the performance of our multifactorial approach ($\text{MF-IPS}^{Mul}$) is much more stable for all values of $\gamma$, and its MSE value closely approximates that of the ground-truth method $\text{MF-IPS}^{GT}$.
When $\gamma < 0.7$, $\text{MF-IPS}^{Mul}$ has a substantially lower MSE than $\text{MF-IPS}^{Pos}$, and when $\gamma > 0.1$ the MSE of $\text{MF-IPS}^{Mul}$ is substantially lower than $\text{MF-IPS}^{Pop}$.
There is an exception when $\gamma > 0.8$, when $\text{MF-IPS}^{Mul}$ is outperformed by $\text{MF-IPS}^{Pos}$ and $\text{MF-IPS}^{GT}$ by a small but noticeable margin.

Similar observations can be made in terms of MAE performance, however, the MAE results have more variance making the trends less clearly apparent.
The increased variance is likely because all methods optimize the MSE in their loss, and thus, they do not necessarily fully minimize the MAE in the process.

Overall, our results show that the performance of single-factor debiasing methods varies greatly depending on how much selection bias is affected by their corresponding factor.
Conversely, the performance of our multifactorial method is hardly affected by how much selection bias depends on each factor, with only showing a minor decrease when selection bias is very close to positivity bias.
Therefore, we answer \textbf{RQ3} in the affirmative:
we conclude that our multifactorial method has the most robust performance and is the safest choice if selection bias could depend on multiple factors.

%% file: sections/sec-rel.tex
\vspace*{-2mm}
\section{Related Work}

Selection bias is pervasive in user interactions with \acp{RS} and can be observed in both explicit feedback (\eg user ratings)~\cite{schnabel2016recommendations,huang2022different} and implicit feedback (\eg user clicks)~\cite{saito2020unbiased,yang2018unbiased}.
As discussed in Section~\ref{sec:bias-conceptualization}, selection bias can arise for a variety of reasons, resulting in different forms of selection bias, such as the well-known popularity bias~\cite{steck2011item,pradel2012ranking,zhu2021popularity,abdollahpouri2020multi} and positivity bias~\cite{pradel2012ranking,park2018positivity}.
Other forms of bias include incentive bias, manifested when users are incentivized to provide ratings for benefits and rewards~\cite{panniello2016impact}, and conformity bias, manifested when users tend to rate items similarly to others in a group~\cite{krishnan2014methodology,knyazev2022bandwagon}.
Such forms of bias are determined by one factor, referred to as single-factor bias.

In reality, selection bias in user interactions can be characterized as a combination of multiple biases or a complex bias that is determined by more than one factor~\cite{zheng2022cbr,wu2021unbiased}.
Previous work suggests that selection
bias is also affected by the additional factor of time~\cite{huang2022different}.
Many contextual factors such as position, modality or surrounding items can result in selection bias in user rating behavior simultaneously~\cite{wu2021unbiased,zhuang2021cross,sarvi-2023-impact}.
Additionally, correlations between selection and both popularity and positivity were observed in multiple real-world datasets~\cite{pradel2012ranking,huang2020keeping}.
Building on this, our focus in this paper is on a multifactorial bias determined by item and rating value factors, which can be seen as a generalization of popularity and positivity.

Debiased recommendation methods aim to mitigate the negative effects of bias and involve both bias estimation and correction~\cite{schnabel2016recommendations,chen2023bias,zhu2021popularity}.
A prevalent family of debiasing methods is based on \acf{IPS}~\cite{imbens2015causal,joachims2017unbiased,schnabel2016recommendations}.
\ac{IPS} weights observations inversely to their observation probability; in theory, its estimation is unbiased but can suffer from high variance~\cite{schnabel2016recommendations}.
Propensity clipping~\cite{saito2020unbiased,chen2019top} and doubly-robust estimation~\cite{wang2019doubly,saito2019doubly,oosterhuis2023doubly} are two common ways to reduce variance for \ac{IPS}.
An alternative research direction involves two-tower methods, which jointly model user-item interactions and estimate bias present in the interactions~\cite{zhuang2021cross}.
Due to a lack of explicit signals of the bias effect on interactions, two-tower methods encounter challenges in distinguishing between user preference modeling and bias estimation~\cite{fedus2022switch}.
In light of this, for our proposed multifactorial method we have chosen to build on the \ac{IPS}-based debiasing method.

Bias or propensity estimation aims to estimate the probability of a user interacting with an item~\cite{schnabel2016recommendations,mccaffrey2004propensity,li2023propensity,zhu2020unbiased,liu2024estimating}.
It is a key component in \ac{IPS} weighting and significantly affects the performance of \ac{IPS} in mitigating bias.
A prevalent method for propensity estimation uses naive Bayes with maximum likelihood, which is commonly used to estimate popularity bias and positivity bias~\cite{schnabel2016recommendations,canamares2018should,zhang2021causal}.
An alternative for propensity estimation is based on optimizing machine learning models.
E.g., logistic regression and \ac{MF} models can be trained to predict propensities that can best generate an observation matrix~\cite{saito2020unbiased,huang2022different,schnabel2016recommendations}.
While  the idea of estimating propensities through optimization is conceptually appealing, our experiments show that these estimates are often unstable and do not always provide propensities that work well with \ac{IPS}.

%% file: sections/sec-con.tex
\section{Conclusion}
We have considered a multifactorial selection bias that is determined by two factors: the item and rating value.
We introduced a propensity estimation method for multifactorial bias and integrated it into the prevalent \ac{IPS}-based debiasing approach.
Furthermore, we proposed the adoption of propensity smoothing and a novel alternating gradient descent method to deal with the sparsity problem that arises in multifactorial bias estimation.

Our experimental results on two real-world datasets show the effectiveness of our multifactorial method over state-of-the-art single-factor counterparts.
Moreover, through a simulation analysis, we found that the performance of our multifactorial method is stable as the effect of different factors is widely varied, in stark contrast with existing single-factor methods.
Thereby, our multifactorial approach appears to be both substantially more robust and significantly effective than previous single-factor debiasing techniques.
Our multifactorial debiasing approach could be an important contribution to the \ac{RS} field, as multifactorial bias appears to better capture real-world forms of bias.

A limitation of our work is that we only consider multifactorial bias in explicit feedback and the rating prediction task.
Future work could extend our multifactorial method to implicit feedback and other recommendation settings, \eg large language models as \acp{RS}.

%% file: sections/acknowledge.tex
\vspace*{-2mm}
\begin{acks}
This research was partially supported by
the Dutch Research Council (NWO)
under pro\-ject numbers, 
024.\-004.\-022,
NWA.1389.20.183,
KICH3.LTP.20.006,
and grant No. VI.Veni.222.\-269,
and by the European Union's Horizon Europe program under No 101070212.

All content represents the opinion of the authors, which is not necessarily shared or endorsed by their respective employers and/or sponsors.
\end{acks}